\documentclass[10pt]{article}
\usepackage{bbm}
\usepackage{graphicx}
\usepackage{amsmath}
\usepackage{amssymb}
\usepackage{caption2}
\begin{document}
\bibliographystyle{prsty}
\begin{center}
 {\large {\bf \sc{ Analysis of the vertexes  $\Omega^*_Q\Omega_Q^*\phi$, $\Omega^*_Q\Xi_Q^*K^*$,
 $\Xi_Q^*\Sigma^*_QK^*$ and $\Sigma_Q^*\Sigma^*_Q \rho$ with the light-cone QCD sum rules }}} \\[2mm]
Zhi-Gang Wang \footnote{E-mail,wangzgyiti@yahoo.com.cn. }, Jun-Fang Li    \\
 Department of Physics, North China Electric Power University, Baoding 071003, P. R. China \\
\end{center}

\begin{abstract}
 In this article, we parameterize the vertexes $\Omega^*_Q\Omega_Q^*\phi$, $\Omega^*_Q\Xi_Q^*K^*$,
 $\Xi_Q^*\Sigma^*_QK^*$ and $\Sigma_Q^*\Sigma^*_Q \rho$ with four
tensor structures due to Lorentz invariance, and study the
 corresponding four strong coupling constants  with the light-cone QCD sum rules.
\end{abstract}


\section{Introduction}
The charmed and bottom baryon states which  contain a heavy quark
and two light quarks are particularly interesting for studying the
dynamics of the light quarks in the presence  of a heavy quark. They
serve as an excellent ground for testing predictions of the quark
models and heavy quark symmetry \cite{ReviewH1,ReviewH2}.  The mass
spectrum and magnetic moments of the heavy (and doubly heavy)
baryon states have been studied by a number of the theoretical
approaches
\cite{H-baryon-1,H-baryon-2,H-baryon-3,H-baryon-4,H-baryon-5,H-baryon-6,H-baryon-7,H-baryon-8,H-baryon-9,
H-baryon-10,H-baryon-11,H-baryon-12,H-baryon-13}. Among those
theoretical approaches,  the QCD sum rules and light-cone QCD sum
rules are powerful  tools in studying the ground state heavy baryons
and have given many successful descriptions of the properties
\cite{SVZ79,PRT85,LCSR89,LCSR,LCSRreview,NarisonBook}.

 In
Refs.\cite{Wang0909,WangEPJA},  we study the strong coupling
constants in the vertexes $\Omega_Q^*\Omega_Q \phi$, $\Xi_Q^*\Xi'_Q
V$ and $\Sigma_Q^*\Sigma_Q V$ with the light-cone QCD sum rules,
then assume the vector meson dominance of the intermediate
$\phi(1020)$, $\rho(770)$ and $\omega(782)$, and calculate the
radiative decays $\Omega_Q^*\to \Omega_Q \gamma$, $\Xi_Q^*\to\Xi'_Q
\gamma$ and  $\Sigma_Q^*\to\Sigma_Q \gamma$. In
Refs.\cite{Lee2010,Azizi2009},  the strong coupling constants
$g_{\Xi_Q\Xi_Q\pi}$ and $g_{\Sigma_{Q}\Lambda_{Q}\pi}$ are
calculated using  the light-cone QCD sum rules. In this article, we
analyze the vertexes $\Omega_Q^*\Omega_Q^*\phi$,
$\Omega_Q^*\Xi_Q^*K^*$, $\Xi_Q^*\Sigma_Q^*K^*$ and
$\Sigma_Q^*\Sigma_Q^*\rho$, which are of great phenomenological
importance, and study the corresponding strong coupling constants
with the light-cone QCD sum rules.

The baryon resonances can be classified  as  genuine $qqq$ states
(or large-$N_c$ ground states) and molecule-like states generated
dynamically, and the two-pole nature of the $\Lambda(1405)$ with
$I(J^P)=0({\frac{1}{2}}^-)$ serves as an excellent support for the
hadrogenesis conjecture \cite{OsetPRL}. The negative-parity baryon
resonances (some are supposed to be molecule-like states) can be
studied through the meson-baryon scatterings  in the coupled-channel
unitary schemes, where the tree-level scattering kernels are derived
from the $SU(3)$ chiral lagrangian \cite{Oset00}, the flavor-spin
$SU(6)$ extension of the Weinberg-Tomozawa meson-baryon interactions
\cite{SU6}, the $t$-channel vector meson exchange model based on the
flavor $SU(4)$ symmetry combined with the chiral symmetry
\cite{SU4-t-1,SU4-t-2}, the flavor-spin $SU(8)$ extension of the
Weinberg-Tomozawa meson-baryon interactions \cite{SU8}, the flavor
$SU(4)$ $t$-channel vector meson exchange model \cite{SU4-t}, etc.
In the limit $t\to 0$,  the vector meson exchange models and the
Weinberg-Tomozawa interactions result in analogous scattering
kernels, where the strong coupling constants in the vertexes $BBV$,
$B^*BV$and $B^*B^*V$ play an important role, for example, the strong
coupling constant in the vertex $\Omega_c^*\Omega_c^*\phi$ for the
$t$-channel $\phi(1020)$-exchange induced scattering $\Omega_c^*+D_s
\to \Omega_c^*+D_s$. In the real world, the flavor $SU(4)$ and the
spin $SU(2)$ are (badly) broken, an universal coupling constant is
not a good approximation, we should calculate those coupling
constants in different channels independently  to estimate the
symmetry breaking effects.

In recent years, the Babar, Belle, CLEO, D0, CDF and FOCUS
collaborations have discovered (or confirmed) a  number of new heavy
baryon states \cite{New-Baryon-1,New-Baryon-2}, and some states have
been studied in the  coupled-channel unitary schemes, for example,
the $\Lambda_c(2595)$ is tentatively identified as a $D^*N$ or $DN$
molecular state \cite{SU8,Mizutani-2006,Lutz-2006}. The
coupled-channel unitary approaches have predicted many heavy baryon
resonances, which maybe discovered at the LHCb, RHIC, PANDA, etc, we
should study the strong coupling constants in the vertexes $BBV$,
$B^*BV$ and $B^*B^*V$ in great details to make the predictions more
reliable.

The article is arranged as follows:  we derive the strong coupling
constants   in the vertexes $B^*B^* V$ with the light-cone QCD sum
rules in Sect.2; in Sect.3, we present the
 numerical results and discussions; and Sect.4 is reserved for our
conclusions.
\section{ The vertexes  $B^*B^* V$  with light-cone QCD sum rules}
We parameterize the vertexes $B^*B^*V$ with four tensor structures
due to Lorentz invariance and introduce four  strong  coupling
constants $g_1$, $g_2$,  $g_3$  and $g_4$ \cite{V1,V2},
\begin{eqnarray}
\langle B^*_i(p+q)|B_j^*(p)V(q) \rangle&=&\overline{U}_i^\alpha(p+q)
\left\{ \left[g_1g_{\alpha\beta}+g_3\frac{q_\alpha
q_\beta}{\left(M_i+M_j\right)^2}\right]\not\!\!{\epsilon}
\right.\nonumber\\
&&\left. -\left[g_2\frac{g_{\alpha\beta}}{M_i+M_j}+g_4\frac{q_\alpha
q_\beta}{\left(M_i+M_j\right)^3}\right]\not\!\!{\epsilon}\not\!\!{q}\right\}U_j^\beta(p)
\, ,
\end{eqnarray}
where the $U^i_\mu(p)$ is the Rarita-Schwinger spinor of the
 heavy baryon states $B^*_i$ ($\Omega_Q^*$,
$\Xi_Q^*$, $\Sigma_Q^*$), the $\epsilon_\mu$ is the polarization
vector of the vector mesons $V$ ($\phi(1020)$, $K^*(892)$,
$\rho(770)$).

In the following, we write down the
 two-point correlation functions  $\Pi_{ijV}(p,q)$,
\begin{eqnarray}
\Pi_{ijV}(p,q)&=&w^\mu z^\nu i \int d^4x \, e^{-i p \cdot x} \,
\langle 0 |T\left\{ J^i_\mu(0)\bar{J}^j_\nu(x)\right\}|V(q)\rangle \, , \\
J^\Omega_\mu(x)&=& \epsilon^{abc}  s^T_a(x)C\gamma_\mu s_b(x)  Q_c(x)  \, ,  \nonumber \\
J^\Xi_\mu(x)&=& \epsilon^{abc}  q^T_a(x)C\gamma_\mu s_b(x)  Q_c(x)  \, ,  \nonumber \\
J^\Sigma_\mu(x)&=& \epsilon^{abc}  q^T_a(x)C\gamma_\mu q'_b(x)
Q_c(x) \, ,
\end{eqnarray}
where $Q=c,b$ and $q,q'=u,d$, the $a,b,c$ are color indexes,  the
Ioffe-type  heavy baryon currents $J^\Omega_\mu(x)$, $J^\Xi_\mu(x)$,
$J^\Sigma_\mu(x)$  interpolate the  $\frac{3}{2}^+$ heavy
 baryon states $\Omega_Q^*$, $\Xi_Q^*$, $\Sigma_Q^*$,  respectively, the external
 vector mesons  have the four momentum $q_\mu$ with $q^2=m_{\phi/K^*/\rho}^2$.

We can insert a complete set  of intermediate hadronic states with
the same quantum numbers as the current operators  $J_\mu(x)$ into
the correlation functions $\Pi_{ijV}(p,q)$  to obtain the hadronic
representation. After isolating the ground state contributions from
the pole terms of the heavy  baryons $\Omega_Q^*$, $\Xi_Q^*$,
$\Sigma_Q^*$, we get the following results,
\begin{eqnarray}
\Pi_{ijV}(p,q)&=&\frac{\langle0| w\cdot J_i(0)|
B^*_i(q+p)\rangle\langle B_i^*(q+p)| B_j^*(p) V(q) \rangle \langle
B_j^*(p)|z \cdot \bar{J}_j(0)| 0\rangle}
  {\left[M_{i}^2-(q+p)^2\right]\left[M_{j}^2-p^2\right]}  + \cdots \nonumber \\
&=& \frac{\lambda_{i}\lambda_{j}}
{\left[M_{i}^2-(q+p)^2\right]\left[M_{j}^2-p^2\right]}
\left[\widetilde{g}_1 \!\not\!{p} p\cdot\epsilon w \cdot z
+\widetilde{g}_2 \!\not\!{p}\!\not\!{q} p\cdot\epsilon w \cdot z
    \right.\nonumber\\
    &&\left.+\widetilde{g}_3 \!\not\!{p} p\cdot\epsilon q\cdot w q\cdot z
+\widetilde{g}_4 \!\not\!{q} p\cdot\epsilon q\cdot w q\cdot
z+\cdots\right]+\cdots \, ,
\end{eqnarray}
where the following definitions have been used,
\begin{eqnarray}
\langle 0| J^i_\mu (0)|B_i^*(p)\rangle &=&\lambda_{i} U^i_\mu(p,s) \, , \nonumber \\
\sum_s U^i_\mu(p,s) \overline{U}^i_\nu(p,s)
&=&-(\!\not\!{p}+M_{i})\left( g_{\mu\nu}-\frac{\gamma_\mu
\gamma_\nu}{3}-\frac{2p_\mu p_\nu}{3M_{i}^2}+\frac{p_\mu
\gamma_\nu-p_\nu \gamma_\mu}{3M_{i}} \right) \,  ,\nonumber
\end{eqnarray}
\begin{eqnarray}
\widetilde{g}_1&=&2g_1 \, , \nonumber \\
\widetilde{g}_2&=&\frac{2g_2}{M_i+M_j} \, , \nonumber \\
\widetilde{g}_3&=&-\frac{4g_1}{3M_i^2}+\frac{4g_2}{3M_i(M_i+M_j)} +\frac{g_3}{(M_i+M_j)^2}\left[2-\frac{2(M_i^2-M_j^2)}{3M_i^2}-\frac{2m_V^2}{3M_i^2} \right]
 \nonumber \\
 &&+\frac{2m_V^2g_4}{3M_i(M_i+M_j)^3}\, ,\nonumber \\
 \widetilde{g}_4&=&-\frac{4g_1}{3M_i^2}-\frac{4g_2}{3M_i(M_i+M_j)} +\frac{g_3}{(M_i+M_j)^2}\left[\frac{4}{3}-\frac{2(M_i^2-M_j^2)}{3M_i^2}-\frac{2m_V^2}{3M_i^2} \right]
 \nonumber \\
 &&+\frac{g_4}{(M_i+M_j)^3}\left[2M_i-\frac{4(M_i^2-M_j^2)}{3M_i}-\frac{2m_V^2}{3M_i}\right]\, .
\end{eqnarray}
In calculation, we have ordered the Dirac matrixes as
$\!\not\!{w}\!\not\!{p}\!\not\!{\epsilon}\!\not\!{q}\!\not\!{z}$
\cite{Aliev0904}.

 The current $J_\mu(x)$ couples
not only to the  spin-parity $J^P=\frac{3}{2}^+$ states, but also to
the   spin-parity $J^P=\frac{1}{2}^-$ states.
 For a generic $\frac{1}{2}^-$ resonance   $B_Q$,
$ \langle0|J_{\mu}(0)|B_Q(p)\rangle=\lambda_{*}
 (\gamma_{\mu}-4\frac{p_{\mu}}{M_{*}})U^{*}(p,s)$, where $\lambda^{*}$ is  the  pole residue, $M_{*}$ is the
mass, and the spinor $U^*(p,s)$  satisfies the usual Dirac equation
$(\not\!\!p-M_{*})U^{*}(p)=0$.
 In this article, we choose the tensor structures $\!\not\!{p} p\cdot \epsilon w \cdot
 z$, $\!\not\!{p} \!\not\!{q} p\cdot \epsilon w \cdot z$, $\!\not\!{p} p\cdot \epsilon q \cdot w q\cdot z$,
 $\!\not\!{q} p\cdot \epsilon q \cdot w q\cdot z$,  the negative-parity baryon state $B_Q$  has no
 contamination.
 For example, we can
study the contribution of the $\frac{1}{2}^-$ baryon state $B_Q$ to
the correlation functions $\Pi_{ijV}(p,q)$,
\begin{eqnarray}
\Pi_{ijV}(p,q)&=&\frac{\langle0| w\cdot J_i(0)|
B_i(q+p)\rangle\langle B_i(q+p)| B_j(p) V(q) \rangle \langle
B_j(p)|z\cdot \bar{J}_j(0)| 0\rangle}
  {\left[M_{*i}^2-(q+p)^2\right]\left[M_{*j}^2-p^2\right]}  + \cdots \nonumber \\
&=&\lambda_{*i}\lambda_{*j} \left[\!\not\!{w}-4\frac{(p+q) \cdot
w}{M_{*i}}\right]\frac{\!\not\!{q}+\!\not\!{p}+M_{*i}}
{M_{*i}^2-(q+p)^2} \left[g_V \!\not\!{\epsilon}
    +ig_T \frac{ \epsilon^\alpha \sigma_{\alpha\beta}
    q^\beta}{M_{*i}+M_{*j}}\right]
    \frac{\!\not\!{p}+M_{*j}}{M_{*j}^2-p^2}
\nonumber\\
&&\left[\!\not\!{z}-4\frac{p \cdot z}{M_{*j}}\right]+\cdots\, ,
\nonumber \\
&=& 0 \!\not\!{p} p\cdot\epsilon w \cdot z +0 \!\not\!{p}\!\not\!{q}
p\cdot\epsilon w \cdot z
    +0 \!\not\!{p} p\cdot\epsilon q\cdot w q\cdot z
+0 \!\not\!{q} p\cdot\epsilon q\cdot w q\cdot z+\cdots \, ,
\end{eqnarray}
where we introduce the strong coupling constants $g_V$ and $g_T$ to
parameterize the vertexes  $\langle B_i(q+p)| B_j(p) V(q) \rangle$,
and order the Dirac matrixes as
$\!\not\!{w}\!\not\!{p}\!\not\!{\epsilon}\!\not\!{q}\!\not\!{z}$.

In the following, we briefly outline the  operator product expansion
for the correlation functions  $\Pi_{ijV }(p,q)$  in perturbative
QCD. The calculations are performed at the large space-like momentum
regions $(q+p)^2\ll 0$  and $p^2\ll 0$, which correspond to the
small light-cone distance $x^2\approx 0$ required by the validity of
the operator product expansion. We contract the quark fields in the
correlation functions $\Pi_{ijV}(p,q)$ with Wick theorem,
\begin{eqnarray}
\Pi_{\Omega_Q^*\Omega^*_Q\phi}(p,q)&=&2i\epsilon^{ijk}\epsilon^{i'j'k'}
\int d^4x e^{-i p\cdot x} S_Q^{kk'}(-x)\left\{  Tr\left[\!\not\!{w}
\langle 0|s_j(0)\bar{s}_{j'}(x)|\phi(q)\rangle \!\not\!{z} CS_{ii'}^T(-x)C\right]\right.\nonumber \\
 &&\left.+ Tr\left[ \!\not\!{w} S_{jj'}(-x)\!\not\!{z} C\langle
 0|s_i(0)\bar{s}_{i'}(x) |\phi(q)\rangle^T C\right] \right\}\,,\nonumber\\
\Pi_{\Omega_Q^*\Xi^*_Q K^*}(p,q)&=&2i\epsilon^{ijk}\epsilon^{i'j'k'}
\int d^4x e^{-i p\cdot x} S_Q^{kk'}(-x)  Tr\left[\!\not\!{w} \langle
0|s_j(0)\bar{u}_{j'}(x) |K^*(q)\rangle \!\not\!{z}CS_{ii'}^T(-x)C\right] \, ,\nonumber \\
 \Pi_{\Xi_Q^*\Sigma^*_Q K^*}(p,q)&=&2i\epsilon^{ijk}\epsilon^{i'j'k'} \int d^4x e^{-i p\cdot
x} S_Q^{kk'}(-x)  Tr\left[\!\not\!{w} \langle 0|s_j(0)\bar{u}_{j'}(x) |K^*(q)\rangle \!\not\!{z}CU_{ii'}^T(-x)C\right] \, ,\nonumber \\
 \Pi_{\Sigma_Q^*\Sigma^*_Q  \rho}(p,q)&=&2i\epsilon^{ijk}\epsilon^{i'j'k'} \int d^4x e^{-i
p\cdot x} S_Q^{kk'}(-x)  Tr\left[\!\not\!{w} \langle
0|d_j(0)\bar{u}_{j'}(x) |\rho(q)\rangle \!\not\!{z} CD_{ii'}^T(-x)C\right] \, ,\nonumber \\
\end{eqnarray}
then perform the  Fierz re-ordering to extract the contributions
from the two-particle  vector meson light-cone distribution
amplitudes, substitute  the full $s$, $u$, $d$  and $Q$ quark
propagators ($S(x)$, $U(x)$, $D(x)$ and $S_Q(x)$)  into the
correlation functions and complete   the integrals  in the
coordinate space and  momentum space sequentially, and obtain the
correlation functions  at the level of quark-gluon degree's of
freedom.  In calculation, the two-particle vector meson light-cone
distribution amplitudes up to twist-4 have been used
\cite{VMLC981,VMLC982,VMLC2003,VMLC2007}. The parameters in the
light-cone distribution amplitudes are scale dependent and are
estimated with the QCD sum rules \cite{VMLC2003,VMLC2007}. In this
article, the energy scale $\mu$ is chosen to be $\mu=1\,\rm{GeV}$.

Taking double Borel transform  with respect to the variables
$Q_1^2=-p^2$ and $Q_2^2=-(p+q)^2$ respectively,  then subtracting
the contributions from the high resonances and continuum states by
introducing  the threshold parameter $s_0$ (i.e. $ M^{2n}\rightarrow
\frac{1}{\Gamma[n]}\int_0^{s_0} ds s^{n-1}e^{-\frac{s}{M^2}}$),
finally we can obtain 32  sum rules  for the strong coupling
constants $\widetilde{g}_{1}$, $\widetilde{g}_{2}$,
$\widetilde{g}_{3}$ and $\widetilde{g}_{4}$ respectively, the
explicit expressions are presented in the Appendix. In calculation,
we neglect the contributions from the high dimension vacuum
condensates, such as $\langle f_{abc}G^aG^bG^c\rangle$, $\langle
\bar{q}q\rangle\langle\frac{\alpha_sGG}{\pi}\rangle$, $\langle
\bar{s}s\rangle\langle\frac{\alpha_sGG}{\pi}\rangle$, etc. They are
greatly suppressed by the large  numerical denominators and
additional inverse powers of the  Borel parameter $\frac{1}{M^{2}}$,
and would not play any significant roles.

\section{Numerical result and discussion}

The parameters which determine the vector meson light-cone
distribution amplitudes are $f_\phi=(0.215\pm0.005)\,\rm{GeV}$,
$f_\phi^{\perp}=(0.186\pm0.009)\,\rm{GeV}$, $a_1^{\parallel}=0.0$,
$a_1^{\perp}=0.0$, $a_2^{\parallel}=0.18\pm0.08$,
$a_2^{\perp}=0.14\pm0.07$, $\zeta^{\parallel}_3=0.024\pm 0.008$,
$\widetilde{\lambda}_3^{\parallel}=0.0$,
$\widetilde{\omega}_3^{\parallel}=-0.045\pm 0.015$,
$\kappa_3^{\parallel}=0.0$, $\omega_3^\parallel=0.09\pm0.03$,
$\lambda_3^\parallel=0.0$, $\kappa_3^\perp=0.0$,
$\omega_3^\perp=0.20\pm0.08$, $\lambda_3^\perp=0.0$,
$\varsigma_4^\parallel=0.00\pm 0.02$,
$\widetilde{\omega}_4^\parallel=-0.02\pm0.01$,
$\varsigma_4^\perp=-0.01\pm 0.03$,
$\widetilde{\varsigma}_4^\perp=-0.03\pm 0.04$,
$\kappa_4^\parallel=0.0$, $\kappa_4^\perp=0.0$ for the $\phi$-meson;
$f_{K^*}=(0.220\pm0.005)\,\rm{GeV}$,
$f_{K^*}^{\perp}=(0.185\pm0.009)\,\rm{GeV}$,
$a_1^{\parallel}=0.03\pm0.02$, $a_1^{\perp}=0.04\pm0.03$,
$a_2^{\parallel}=0.11\pm0.09$, $a_2^{\perp}=0.10\pm0.08$,
$\zeta^{\parallel}_3=0.023\pm 0.008$,
$\widetilde{\lambda}_3^{\parallel}=0.035\pm0.015$,
$\widetilde{\omega}_3^{\parallel}=-0.07\pm 0.03$,
$\kappa_3^{\parallel}=0.000\pm0.001$,
$\omega_3^\parallel=0.10\pm0.04$,
$\lambda_3^\parallel=-0.008\pm0.004$,
$\kappa_3^\perp=0.003\pm0.003$, $\omega_3^\perp=0.3\pm0.1$,
$\lambda_3^\perp=-0.025\pm0.020$, $\varsigma_4^\parallel=0.02\pm
0.02$, $\widetilde{\omega}_4^\parallel=-0.02\pm0.01$,
$\varsigma_4^\perp=-0.01\pm 0.03$,
$\widetilde{\varsigma}_4^\perp=-0.05\pm 0.04$,
$\kappa_4^\parallel=-0.025\pm0.005$, $\kappa_4^\perp=0.013\pm0.005$
for the $K^*$-meson; and $f_\rho=(0.216\pm0.003)\,\rm{GeV}$,
$f_\rho^{\perp}=(0.165\pm0.009)\,\rm{GeV}$, $a_1^{\parallel}=0.0$,
$a_1^{\perp}=0.0$, $a_2^{\parallel}=0.15\pm0.07$,
$a_2^{\perp}=0.14\pm0.06$, $\zeta^{\parallel}_3=0.030\pm 0.010$,
$\widetilde{\lambda}_3^{\parallel}=0.0$,
$\widetilde{\omega}_3^{\parallel}=-0.09\pm 0.03$,
$\kappa_3^{\parallel}=0.0$, $\omega_3^\parallel=0.15\pm0.05$,
$\lambda_3^\parallel=0.0$, $\kappa_3^\perp=0.0$,
$\omega_3^\perp=0.55\pm0.25$, $\lambda_3^\perp=0.0$,
$\varsigma_4^\parallel=0.07\pm 0.03$,
$\widetilde{\omega}_4^\parallel=-0.03\pm0.01$,
$\varsigma_4^\perp=-0.03\pm 0.05$,
$\widetilde{\varsigma}_4^\perp=-0.08\pm 0.05$,
$\kappa_4^\parallel=0.0$, and $\kappa_4^\perp=0.0$ for the
$\rho$-meson  at the energy scale $\mu=1\, \rm{GeV}$
\cite{VMLC2003,VMLC2007}.

The QCD input parameters are taken to be the standard values
$m_s=(140\pm 10 )\,{\rm{MeV}}=24.6\,m_{u/d}$, $m_c=(1.35\pm
0.10)\,\rm{GeV}$, $m_b=(4.7\pm 0.1)\,\rm{GeV}$,
 $\langle \bar{q}q \rangle=-(0.24\pm
0.01 \,\rm{GeV})^3$, $\langle \bar{s}s \rangle=(0.8\pm 0.2 )\langle
\bar{q}q \rangle$, $\langle \bar{s}g_s\sigma Gs \rangle=m_0^2\langle
\bar{s}s \rangle$, $\langle \bar{q}g_s\sigma Gq \rangle=m_0^2\langle
\bar{q}q \rangle$, $m_0^2=(0.8 \pm 0.2)\,\rm{GeV}^2$, and $\langle
\frac{\alpha_s GG}{\pi}\rangle=(0.33\,\rm{GeV})^4 $  at the energy
scale $\mu=1\, \rm{GeV}$ \cite{SVZ79,PRT85,Ioffe2005}.

The masses of the well established  hadrons are taken from the
Review of Particle Physics,  $m_\phi=1.019\,\rm{GeV}$,
$m_{K^*}=0.892\,\rm{GeV}$, $m_\rho=0.775\,\rm{GeV}$,
$M_{\Omega_c^*}=2.766\,\rm{GeV}$, $M_{\Xi_c^*}=2.646\,\rm{GeV}$,
$M_{\Sigma_c^{*}}=2.518\,\rm{GeV}$ and
$M_{\Sigma_b^{*}}=5.833\,\rm{GeV}$ \cite{PDG}. The bottom baryon
states $\Xi^*_b$ and $\Omega_b^*$ have not been observed yet, we use
the values from the conventional QCD sum rules,
$M_{\Xi_b^{*}}=6.02\,\rm{GeV}$ and $M_{\Omega_b^{*}}=6.17\,\rm{GeV}$
\cite{Wang32}.  The values of the pole residues $\lambda_i$ are also
determined   with the conventional QCD sum rules,
$\lambda_{\Omega^*_b}=(0.083\pm0.018)\,\rm{GeV}^3$,
$\lambda_{\Xi^*_b}=(0.049\pm0.012)\,\rm{GeV}^3$,
$\lambda_{\Sigma^*_b}=(0.038\pm0.011)\,\rm{GeV}^3$,
$\lambda_{\Omega^*_c}=(0.056\pm0.012)\,\rm{GeV}^3$,
$\lambda_{\Xi^*_c}=(0.033\pm0.008)\,\rm{GeV}^3$,
$\lambda_{\Sigma^*_c}=(0.027\pm0.008)\,\rm{GeV}^3$  \cite{Wang32}.
In the channels $\Sigma^*_{Qqq}$, we can make a simple replacement
$\lambda_{\Sigma_Q^*} \to \sqrt{2}\lambda_{\Sigma_Q^*}$ to take into
account the symmetry factor.

The threshold parameters $s_0$ and Borel parameters $M^2$  are taken
as $s_0=(12.0\pm1.0)\,\rm{GeV}^2$, $(12.0\pm1.0)\,\rm{GeV}^2$,
$(11.5\pm1.0)\,\rm{GeV}^2$, $(11.0\pm1.0)\,\rm{GeV}^2$,
$(47.0\pm1.0)\,\rm{GeV}^2$, $(47.0\pm1.0)\,\rm{GeV}^2$,
$(46.0\pm1.0)\,\rm{GeV}^2$, $(45.0\pm1.0)\,\rm{GeV}^2$ and
$M^2=(2.4-3.4)\,\rm{GeV}^2$, $(2.3-3.3)\,\rm{GeV}^2$,
$(2.1-3.1)\,\rm{GeV}^2$, $(2.0-3.0)\,\rm{GeV}^2$,
$(5.3-6.3)\,\rm{GeV}^2$, $(5.2-6.2)\,\rm{GeV}^2$,
$(4.8-5.8)\,\rm{GeV}^2$, $(4.6-5.6)\,\rm{GeV}^2$ for the strong
coupling constants in the vertexes $\Omega_c^* \Omega_c^*\phi $,
$\Omega_c^*\Xi_c^* K^* $, $\Xi_c^*\Sigma_c^*K^* $, $\Sigma_c^*
\Sigma_c^*\rho $, $\Omega_b^* \Omega_b^*\phi $, $\Omega_b^*
\Xi_b^*K^* $, $\Xi_b^*\Sigma_b^* K^*$, $\Sigma_b^* \Sigma_b^*\rho $,
respectively \cite{Wang32}.

In Ref.\cite{Wang32}, we study the masses and pole residues of the
${\frac{3}{2}}^+$ heavy baryon states with the conventional
two-point QCD sum rules, and obtain the optimal Borel parameters
$M^2$ and threshold parameters $s_0$. The central values of the
threshold parameters are $s_0\approx(M_{B^*_i}+0.7\,\rm{GeV})^2$,
which can take into account the ground state contributions
sufficiently.  In the constituent quark models,  the energy gap
between the ground states and the first radial excited states is
about $0.5\,\rm{GeV}$, the contributions from  the high resonances
and continuum  states may be included in. The values of the strong
coupling constants $g_{1}$, $g_{2}$, $\widetilde{g}_{3}$ and
$\widetilde{g}_{4}$ are not sensitive to the threshold parameters,
the contaminations should  be very small due to the suppression
factor $e^{-x} < 4\%$ and $ \ll 1\%$ in  the charmed and bottom
channels respectively, where $x=\frac{s_0}{M^2}$. In Fig.1, we plot
the values of the strong coupling constants $g_1$ with variation of
the threshold parameter $s_0$ for the central values of the other
parameters to illustrate the fact. On the other hand,  the values of
the strong coupling constants $g_{1}$, $g_{2}$, $\widetilde{g}_{3}$
and $\widetilde{g}_{4}$ are rather stable with variations of the
Borel parameter, the uncertainties originate from the Borel
parameters are not large. In Fig.2, we plot the values of the strong
coupling constants $g_1$ with variation of the Borel parameter $M^2$
as an example. The Borel parameters and threshold parameters
determined in Ref.\cite{Wang32} still work  in the present case.

\begin{figure}
 \centering
    \includegraphics[totalheight=5cm,width=7cm]{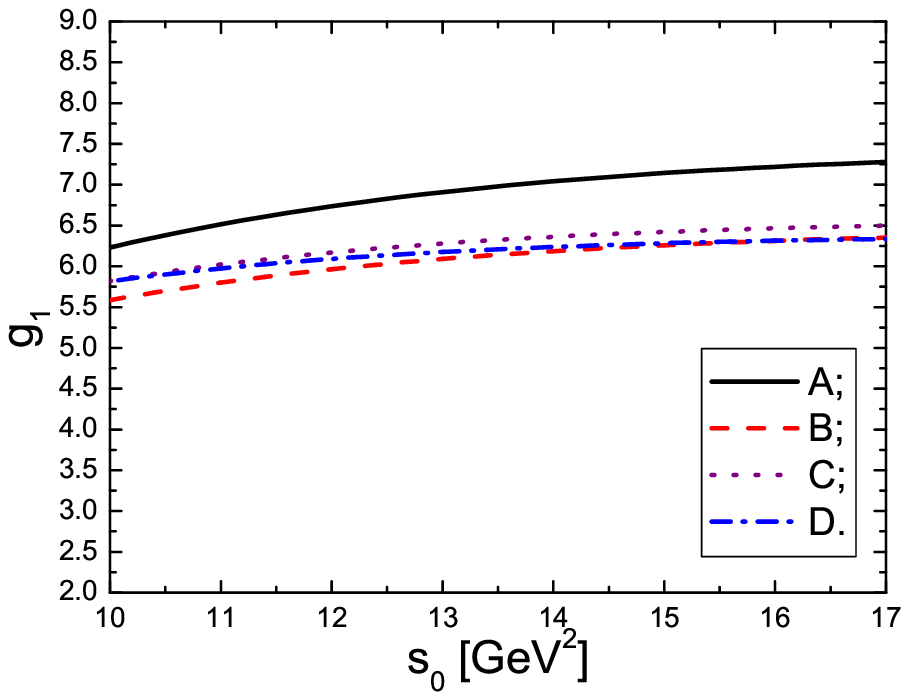}
    \includegraphics[totalheight=5cm,width=7cm]{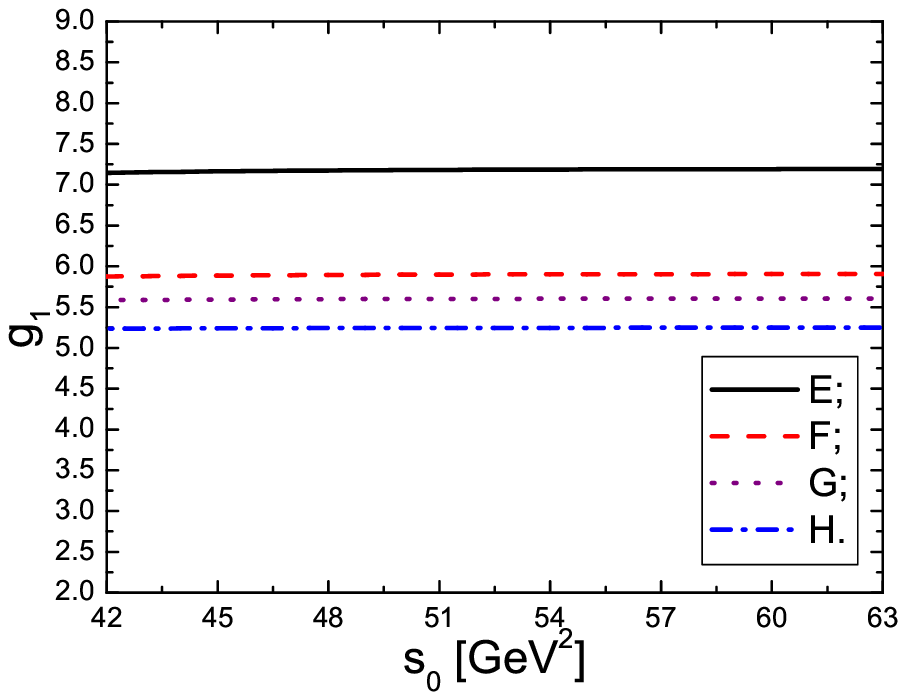}
      \caption{ The values of the strong coupling constants $g_1$ with variation of the threshold parameter
   $s_0$ for the central values of the other parameters. The $A$, $B$, $C$, $D$, $E$, $F$, $G$ and $H$ denote the
   vertexes  $\Omega^*_c\Omega_c^*\phi$, $\Omega^*_c\Xi_c^*K^*$,
 $\Xi_c^*\Sigma^*_cK^*$, $\Sigma_c^*\Sigma^*_c \rho$, $\Omega^*_b\Omega_b^*\phi$, $\Omega^*_b\Xi_b^*K^*$,
 $\Xi_b^*\Sigma^*_bK^*$ and $\Sigma_b^*\Sigma^*_b \rho$, respectively.}
\end{figure}

\begin{figure}
 \centering
    \includegraphics[totalheight=5cm,width=7cm]{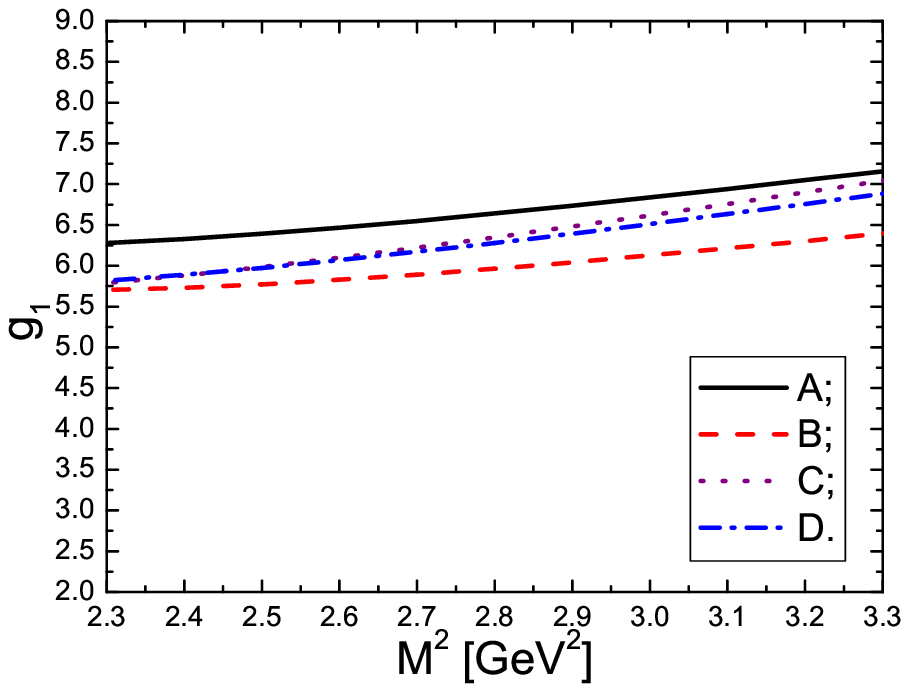}
    \includegraphics[totalheight=5cm,width=7cm]{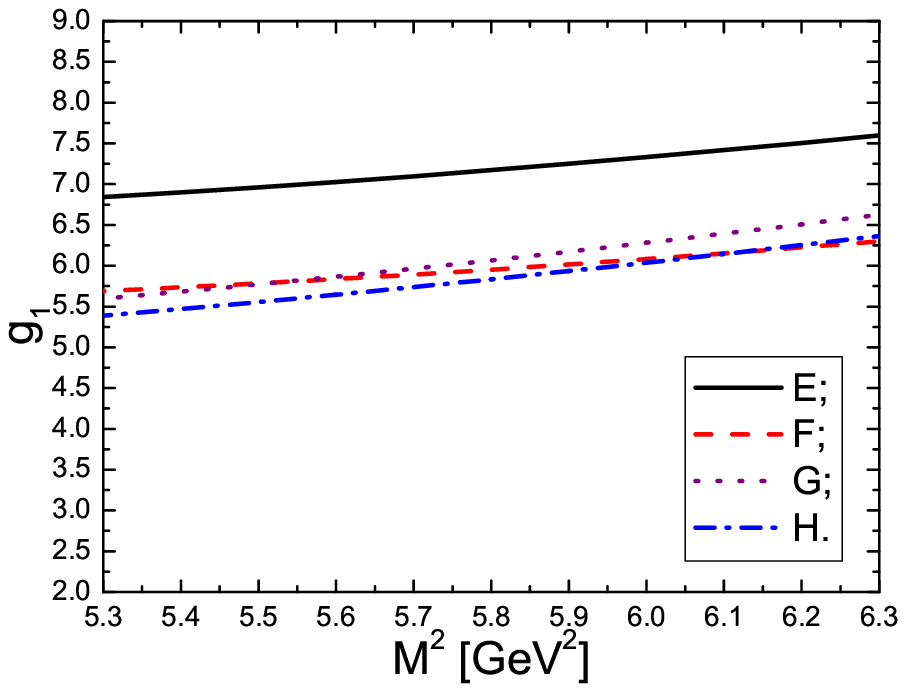}
      \caption{ The values of the strong coupling constants $g_1$ with variation of the Borel parameter
   $M^2$ for the central values of the input parameters. The $A$, $B$, $C$, $D$, $E$, $F$, $G$ and $H$ denote the
   vertexes  $\Omega^*_c\Omega_c^*\phi$, $\Omega^*_c\Xi_c^*K^*$,
 $\Xi_c^*\Sigma^*_cK^*$, $\Sigma_c^*\Sigma^*_c \rho$, $\Omega^*_b\Omega_b^*\phi$, $\Omega^*_b\Xi_b^*K^*$,
 $\Xi_b^*\Sigma^*_bK^*$ and $\Sigma_b^*\Sigma^*_b \rho$, respectively.}
\end{figure}

\begin{table}
\begin{center}
\begin{tabular}{|c|c|c|c|c|}
 \hline\hline Vertexes &$g_1 $&$-g_2 $&$-\widetilde{g}_3(\rm{GeV}^{-2})$&$-\widetilde{g}_{4}(\rm{GeV}^{-2})$ \\
 \hline $\Omega_c^* \Omega_c^*\phi $&  $6.80^{+3.84}_{-2.85}$ & $0.036^{+0.035}_{-0.020}$ & $2.50^{+2.06}_{-1.21}$ & $1.24^{+1.03}_{-0.60}$ \\
 \hline $\Omega_c^* \Xi_c^*K^* $& $6.05^{+3.45}_{-2.46}$ & $0.034^{+0.034}_{-0.019}$ & $9.21^{+8.68}_{-5.06}$ &$4.59^{+4.33}_{-2.52}$ \\
 \hline $\Xi_c^*\Sigma_c^* K^* $& $6.20^{+4.54}_{-2.88}$  & $0.044^{+0.052}_{-0.026}$ & $11.94^{+12.78}_{-6.81}$ & $5.96^{+6.36}_{-3.40}$ \\
 \hline $\Sigma_c^* \Sigma_c^*\rho $& $6.11^{+4.55}_{-2.71}$  & $0.053^{+0.068}_{-0.033}$ & $1.39^{+1.56}_{-0.76}$ & $0.68^{+0.78}_{-0.36}$ \\
 \hline $\Omega_b^* \Omega_b^*\phi$&  $7.22^{+4.09}_{-3.00}$ & $0.035^{+0.028}_{-0.017}$ & $1.05^{+0.74}_{-0.46}$ & $0.52^{+0.38}_{-0.22}$ \\
 \hline $\Omega_b^* \Xi_b^*K^* $&  $5.94^{+3.47}_{-2.45}$ & $0.029^{+0.025}_{-0.014}$ & $3.62^{+2.90}_{-1.79}$ & $1.81^{+1.47}_{-0.92}$ \\
 \hline $\Xi_b^*\Sigma_b^* K^* $& $5.65^{+4.07}_{-2.62}$  & $0.037^{+0.035}_{-0.019}$ & $4.41^{+3.94}_{-2.26}$ & $2.20^{+1.97}_{-1.12}$ \\
 \hline $\Sigma_b^* \Sigma_b^*\rho $&  $5.30^{+3.95}_{-2.40}$ & $0.044^{+0.044}_{-0.023}$  & $0.48^{+0.43}_{-0.23}$ & $0.24^{+0.22}_{-0.11}$ \\
 \hline $\widehat{\Omega_c^* \Omega_c^*\phi} $&  $6.80^{+2.33}_{-2.25}$ & $0.036^{+0.022}_{-0.016}$ & $2.50^{+1.26}_{-0.93}$ & $1.24^{+0.63}_{-0.46}$ \\
 \hline $\widehat{\Omega_c^* \Xi_c^*K^*} $&  $6.05^{+1.89}_{-1.84}$ & $0.034^{+0.020}_{-0.016}$ & $9.21^{+6.05}_{-4.33}$ &$4.59^{+3.02}_{-2.15}$ \\
 \hline $\widehat{\Xi_c^*\Sigma_c^* K^*} $& $6.20^{+2.32}_{-2.14}$  & $0.044^{+0.028}_{-0.021}$ & $11.94^{+8.02}_{-5.64}$ & $5.96^{+3.99}_{-2.82}$ \\
 \hline $\widehat{\Sigma_c^* \Sigma_c^*\rho} $& $6.11^{+1.86}_{-1.79}$  & $0.053^{+0.035}_{-0.026}$ & $1.39^{+0.80}_{-0.55}$ & $0.68^{+0.40}_{-0.26}$ \\
 \hline $\widehat{\Omega_b^* \Omega_b^*\phi}$&  $7.22^{+2.60}_{-2.35}$ & $0.035^{+0.018}_{-0.013}$ & $1.05^{+0.47}_{-0.34}$ & $0.52^{+0.24}_{-0.16}$ \\
 \hline $\widehat{\Omega_b^* \Xi_b^*K^*} $&  $5.94^{+2.04}_{-1.84}$ & $0.029^{+0.016}_{-0.010}$ & $3.62^{+2.00}_{-1.47}$ & $1.81^{+1.04}_{-0.77}$ \\
 \hline $\widehat{\Xi_b^* \Sigma_b^* K^*} $& $5.65^{+2.25}_{-1.97}$  & $0.037^{+0.020}_{-0.014}$ & $4.41^{+2.49}_{-1.79}$ & $2.20^{+1.24}_{-0.89}$ \\
 \hline $\widehat{\Sigma_b^* \Sigma_b^*\rho} $&  $5.30^{+1.91}_{-1.67}$ & $0.044^{+0.023}_{-0.017}$  & $0.48^{+0.23}_{-0.16}$ & $0.24^{+0.11}_{-0.08}$ \\
\hline Average Values &  $6.16$ & $0.039$  &  &  \\
         \hline \hline
\end{tabular}
\end{center}
\caption{ The values of the strong coupling constants $g_1$, $g_2$,
$\widetilde{g}_3$ and $\widetilde{g}_4$, the wide-hat
$\,\,\,\widehat{}\,\,\,$ denotes  the uncertainties originate from
the parameters $\lambda_{i}$ are subtracted.}
\end{table}

Taking into account all the uncertainties of the revelent
parameters, finally we obtain the numerical results of the strong
coupling constants $g_{1}$, $g_{2}$, $\widetilde{g}_{3}$ and
$\widetilde{g}_{4}$, which are shown in the Table 1. We estimate the
uncertainties $\delta$  with the formula
$\delta=\sqrt{\sum_i\left(\frac{\partial f}{\partial
x_i}\right)^2\mid_{x_i=\bar{x}_i} (x_i-\bar{x}_i)^2}$,
 where the $f$ denotes  strong coupling constants $g_{1}$, $g_{2}$, $\widetilde{g}_{3}$ and
$\widetilde{g}_{4}$,  the $x_i$ denotes  the revelent  parameters
$m_Q$, $\langle \bar{q}q \rangle$, $\langle \bar{s}s \rangle$,
$\cdots$. In the numerical calculations, we take the  approximation
$\left(\frac{\partial f}{\partial x_i}\right)^2
(x_i-\bar{x}_i)^2\approx \left[f(\bar{x}_i\pm \Delta
x_i)-f(\bar{x}_i)\right]^2$ for simplicity.  For the central values
of the strong coupling constants,
$\frac{|g_1-\overline{g}_1|}{\overline{g}_1}< 20\%$,
$\frac{|g_2-\overline{g}_2|}{\overline{g}_2}< 40\%$, the heavy quark
symmetry and the light-flavor $SU(3)$ symmetry work rather well.

From Table 1, we can see that the values of the $\widetilde{g}_{3}$
and $\widetilde{g}_{4}$ differ from each other greatly in different
channels, it is no use to obtain an average. In the sum rules for
the strong coupling constants $\widetilde{g}_{3}$ and
$\widetilde{g}_{4}$, the dominant contributions come from the
two-particle twist-4 light-cone distribution amplitude
$\widetilde{A}(1-u_0)$, the light-flavor $SU(3)$ symmetry breaking
effects $m_s\pm m_q$ are too large, i.e.
\begin{eqnarray}
 \widetilde{\widetilde{A}}(1-u_0)=0.093,\,\,\,\widetilde{A}(1-u_0)=0.495, \,\,\, A(1-u_0)=1.445\,  ,\nonumber \\
 \widetilde{\widetilde{A}}(1-u_0)=1.121,\,\,\,\widetilde{A}(1-u_0)=3.356, \,\,\, A(1-u_0)=1.581\,  ,\nonumber \\
 \widetilde{\widetilde{A}}(1-u_0)=0.089,\,\,\,\widetilde{A}(1-u_0)=0.500, \,\,\, A(1-u_0)=1.554\, ,
\end{eqnarray}
for the vector mesons $\phi(1020)$, $K^*(892)$, $\rho(770)$,
respectively; one can consult Ref.\cite{VMLC2007} for the lengthy
expressions of the twist-4 light-cone distribution amplitude $A(u)$.

 The main uncertainties originate
from the parameters $\lambda_{i}$ ($g_1$, $g_2$, $\widetilde{g}_3$
and $\widetilde{g}_4$ $\propto {1\over\lambda_{i} \lambda_{j}}$) and
$m_{Q}$, the variations of those parameters can lead to relatively
large changes for the numerical values, and almost saturate the
total uncertainties, i.e. the variations of the two hadronic
parameters $\lambda_{i}$ and $\lambda_{j}$ lead to an uncertainty
about $25\%\times \sqrt{2}=35\%$, and the variations of the   $m_Q$
lead to an uncertainty about $(10-20)\%$, refining those parameters
is of great importance.  In Table 1, we also present the values of
the strong coupling constants $g_1$, $g_2$, $\widetilde{g}_3$ and
$\widetilde{g}_4$ with the uncertainties originate from the
parameters $\lambda_{i}$ are subtracted. On the other hand, although
there are many parameters in the light-cone distribution amplitudes
\cite{VMLC2003,VMLC2007}, the uncertainties originate from those
parameters are rather small.

Those strong coupling constants in the vertexes $B^*B^*V$ are basic
parameters in describing the interactions among the heavy mesons and
heavy baryons, once reasonable values are obtained, we can use them
to study the meson-baryon scatterings and  perform other
phenomenological analysis. In the present case, the values of the
$g_1$ and $g_2$ are rather good, while the values of the
$\widetilde{g}_3$ and $\widetilde{g}_4$ are not satisfactory, the
two-particle twist-4 light-cone distribution amplitude $A(u)$ should
be re-estimated.

\section{Conclusion}

The strong coupling constants in the vertexes $B^*B^*V$ are basic
parameters in describing the interactions among the heavy mesons and
heavy baryons, where the flavor $SU(4)$ symmetry and the spin
$SU(2)$ symmetry (or the flavor-spin $SU(8)$ symmetry) are badly
broken, an universal coupling constant is not a good approximation.
In this article, we parameterize the vertexes
$\Omega^*_Q\Omega_Q^*\phi$, $\Omega^*_Q\Xi_Q^*K^*$,
$\Xi_Q^*\Sigma^*_QK^*$ and
 $\Sigma_Q^*\Sigma^*_Q \rho$ with four
tensor structures due to Lorentz invariance, study the corresponding
four  strong coupling constants $g_1$, $g_2$, $\widetilde{g}_3$ and
$\widetilde{g}_4$ with the light-cone QCD sum rules. In calculation,
we order  the Dirac matrixes as
$\!\not\!{w}\!\not\!{p}\!\not\!{\epsilon}\!\not\!{q}\!\not\!{z}$ and
choose the tensor structures $\!\not\!{p} p\cdot \epsilon w \cdot
 z$, $\!\not\!{p} \!\not\!{q} p\cdot \epsilon w \cdot z$, $\!\not\!{p} p\cdot \epsilon q \cdot w q\cdot z$,
 $\!\not\!{q} p\cdot \epsilon q \cdot w q\cdot z$ to avoid the contaminations from the
negative-parity heavy baryons states as the interpolating currents
couple to both the  spin-parity $J^P=\frac{3}{2}^+$ and
$J^P=\frac{1}{2}^-$ states.  The final numerical results indicate
that the heavy quark symmetry and the light-flavor $SU(3)$ symmetry
work  rather well for the strong coupling constants $g_1$ and $g_2$,
while the values of the $\widetilde{g}_3$ and $\widetilde{g}_4$
differ from each other greatly in different channels. The dominant
contributions to the strong coupling constants $\widetilde{g}_{3}$
and $\widetilde{g}_{4}$ come from the two-particle twist-4
light-cone distribution amplitude $\widetilde{A}(1-u_0)$, where the
light-flavor $SU(3)$ symmetry breaking effects  are too large and
should be re-estimated.  We can use the strong coupling constants
$g_1$ and $g_2$ to study the dynamically generated molecule-like
states via the meson-baryon scatterings  or perform other
phenomenological analysis.

\section*{Acknowledgment}
This  work is supported by National Natural Science Foundation,
Grant Numbers 10775051, 11075053, and Program for New Century
Excellent Talents in University, Grant Number NCET-07-0282, and the
Fundamental Research Funds for the Central Universities.

\section*{Appendix}

 The  32 sum rules for the strong coupling constants $\widetilde{g}_1$, $\widetilde{g}_2$, $\widetilde{g}_3$
and $\widetilde{g}_4$ in different channels,
\begin{eqnarray}
\widetilde{g}_{\Omega^*_Q\Omega^*_Q\phi}^1&=&\frac{1}{\lambda_{\Omega^*_Q}^2
}\exp{\left[\frac{M_{\Omega^*_Q}^2-u_0\bar{u}_0m_\phi^2}{M^2}\right]}\left\{\frac{f_\phi
m_\phi
\phi_{\parallel}(\bar{u}_0)M^4E_1(x)}{2\pi^2}\int_0^1 dt t\bar{t} e^{-\frac{\widetilde{m}_Q^2}{M^2}}\right.\nonumber\\
 &&- \frac{f_\phi m_\phi m_Q^2
\phi_{\parallel}(\bar{u}_0)}{36M^2}\langle\frac{\alpha_sGG}{\pi}\rangle\int_0^1
dt \frac{\bar{t}}{t^2} e^{-\frac{\widetilde{m}_Q^2}{M^2}}
-\frac{f_\phi m_\phi^3 \widetilde{A}(\bar{u}_0)M^2E_0(x) }{8\pi^2}
\int_0^1 dt t e^{-\frac{\widetilde{m}_Q^2}{M^2}}
\nonumber\\
&&+\frac{f_\phi m_\phi^3m_Q^2 \widetilde{A}(\bar{u}_0)
}{144M^4}\langle\frac{\alpha_sGG}{\pi}\rangle \int_0^1 dt
\frac{1}{t^2} e^{-\frac{\widetilde{m}_Q^2}{M^2}}\nonumber\\
&&-\frac{f_\phi m_\phi^3 \widetilde{\widetilde{C}}(\bar{u}_0)
}{\pi^2} \int_0^1 dt t\bar{t}\left[2M^2E_0(x)+\widetilde{m}_Q^2\right] e^{-\frac{\widetilde{m}_Q^2}{M^2}}\nonumber\\
 &&-\frac{f_\phi m_\phi^3m_Q^2 \widetilde{\widetilde{C}}(\bar{u}_0)
}{18M^4}\langle\frac{\alpha_sGG}{\pi}\rangle \int_0^1 dt \frac{\bar{t}}{t^2}\left[1-\frac{\widetilde{m}_Q^2}{M^2}\right] e^{-\frac{\widetilde{m}_Q^2}{M^2}}\nonumber\\
&&  +\frac{\widetilde{f}^{\perp}_\phi m_\phi^2
m_sh_s^{\parallel}(\bar{u}_0)M^2E_0(x)}{2\pi^2} \int_0^1 dt
t e^{-\frac{\widetilde{m}_Q^2}{M^2}}   \nonumber\\
&&\left.-\frac{\widetilde{f}^{\perp}_\phi m_\phi^2
m_sm_Q^2h_s^{\parallel}(\bar{u}_0)}{36M^4}
\langle\frac{\alpha_sGG}{\pi}\rangle\int_0^1 dt \frac{1}{t^2}
e^{-\frac{\widetilde{m}_Q^2}{M^2}}  +\frac{f_\phi m_\phi
\phi_{\parallel}(\bar{u}_0)}{72}
\langle\frac{\alpha_sGG}{\pi}\rangle\int_0^1 dt
e^{-\frac{\widetilde{m}_Q^2}{M^2}}   \right\} \nonumber\\
&+&\frac{1}{\lambda_{\Omega^*_Q}^2}\exp{\left[\frac{M_{\Omega^*_Q}^2-m_Q^2-u_0\bar{u}_0m_\phi^2}{M^2}\right]}\left\{
\frac{\langle\bar{s}s\rangle f_\phi m_\phi m_s
\phi_{\parallel}(\bar{u}_0)}{3}- \frac{2\langle\bar{s}s\rangle
f_\phi m_\phi^3 m_sm_Q^2
\widetilde{\widetilde{C}}(\bar{u}_0)}{3M^4} \right.\nonumber\\
 &&- \frac{\langle \bar{s}g_s \sigma
Gs\rangle f_\phi m_\phi m_s
g_\perp^{(v)}(\bar{u}_0)}{18M^2}\left(1+\frac{m_Q^2}{M^2}\right)-
\frac{2 \widetilde{f}_\phi^{\perp} m_\phi^2 \langle \bar{s}s\rangle
h_{s}^{\parallel}(\bar{u}_0)}{3}\nonumber\\
 &&\left.+ \frac{\langle \bar{s}g_s \sigma
Gs\rangle f_\phi m_\phi^2
h_{s}^{\parallel}(\bar{u}_0)}{6M^2}\left(1+\frac{m_Q^2}{M^2}\right)
-\frac{ f_\phi m_\phi^3
 \widetilde{\widetilde{C}}(\bar{u}_0)}{36M^2} \langle\frac{\alpha_sGG}{\pi}\rangle +\frac{ f_\phi m_\phi^3
 \widetilde{A}(\bar{u}_0)}{288M^2} \langle\frac{\alpha_sGG}{\pi}\rangle\right\} \,
 , \nonumber\\
 \end{eqnarray}

\begin{eqnarray}
\widetilde{g}_{\Omega^*_Q\Omega^*_Q\phi}^2&=&-\frac{1}{\lambda_{\Omega^*_Q}^2
}\exp{\left[\frac{M_{\Omega^*_Q}^2-u_0\bar{u}_0m_\phi^2}{M^2}\right]}\frac{f_\phi^\perp
m_\phi
m_Qg_{\perp}^{(a)}(\bar{u}_0)}{144M^2}\langle\frac{\alpha_sGG}{\pi}\rangle\int_0^1
dt \frac{1}{t} e^{-\frac{\widetilde{m}_Q^2}{M^2}} \,
 ,
 \end{eqnarray}

\begin{eqnarray}
\widetilde{g}_{\Omega^*_Q\Omega^*_Q\phi}^3&=&\frac{1}{\lambda_{\Omega^*_Q}^2
}\exp{\left[\frac{M_{\Omega^*_Q}^2-u_0\bar{u}_0m_\phi^2}{M^2}\right]}\left\{\frac{2u_0f_\phi
m_\phi
\left[\widetilde{\phi}_{\parallel}(\bar{u}_0)-\widetilde{g}_{\perp}^{(v)}(\bar{u}_0)\right]M^2E_0(x)}{\pi^2}\int_0^1 dt t\bar{t} e^{-\frac{\widetilde{m}_Q^2}{M^2}}\right.\nonumber\\
 &&- \frac{u_0f_\phi m_\phi m_Q^2
\left[\widetilde{\phi}_{\parallel}(\bar{u}_0)-\widetilde{g}_{\perp}^{(v)}(\bar{u}_0)\right]}{9M^4}
\langle\frac{\alpha_sGG}{\pi}\rangle\int_0^1
dt \frac{\bar{t}}{t^2} e^{-\frac{\widetilde{m}_Q^2}{M^2}}\nonumber\\
&&-\frac{u_0f_\phi m_\phi^3 \widetilde{A}(\bar{u}_0) }{2\pi^2}
\int_0^1 dt t e^{-\frac{\widetilde{m}_Q^2}{M^2}}+\frac{u_0f_\phi
m_\phi^3m_Q^2 \widetilde{A}(\bar{u}_0)
}{36M^6}\langle\frac{\alpha_sGG}{\pi}\rangle \int_0^1 dt
\frac{1}{t^2} e^{-\frac{\widetilde{m}_Q^2}{M^2}}\nonumber\\
&&+\frac{2u_0^2f_\phi m_\phi^3 \widetilde{\widetilde{C}}(\bar{u}_0)
}{\pi^2} \int_0^1 dt t\bar{t}
e^{-\frac{\widetilde{m}_Q^2}{M^2}}-\frac{u_0^2f_\phi m_\phi^3m_Q^2
\widetilde{\widetilde{C}}(\bar{u}_0)
}{9M^6}\langle\frac{\alpha_sGG}{\pi}\rangle \int_0^1 dt \frac{\bar{t}}{t^2} e^{-\frac{\widetilde{m}_Q^2}{M^2}}\nonumber\\
&&\left.  +\frac{u_0f_\phi m_\phi
\left[\widetilde{\phi}_{\parallel}(\bar{u}_0)-\widetilde{g}_{\perp}^{(v)}(\bar{u}_0)\right]}{18M^2}
\langle\frac{\alpha_sGG}{\pi}\rangle\int_0^1 dt
e^{-\frac{\widetilde{m}_Q^2}{M^2}}   \right\} \nonumber\\
&+&\frac{1}{\lambda_{\Omega^*_Q}^2}\exp{\left[\frac{M_{\Omega^*_Q}^2-m_Q^2-u_0\bar{u}_0m_\phi^2}{M^2}\right]}\left\{
\frac{4u_0\langle\bar{s}s\rangle f_\phi m_\phi m_s
\left[\widetilde{\phi}_{\parallel}(\bar{u}_0)-\widetilde{g}_{\perp}^{(v)}(\bar{u}_0)\right]}{3M^2}
 \right.\nonumber\\
  &&\left.+\frac{4u^2\langle\bar{s}s\rangle f_\phi m_\phi^3 m_s
\widetilde{\widetilde{C}}(\bar{u}_0)}{3M^4} -\frac{ u_0f_\phi
m_\phi^3
 \widetilde{A}(\bar{u}_0)}{72M^4} \langle\frac{\alpha_sGG}{\pi}\rangle\right\} \,
 ,
 \end{eqnarray}

 \begin{eqnarray}
\widetilde{g}_{\Omega^*_Q\Omega^*_Q\phi}^4&=&\frac{1}{\lambda_{\Omega^*_Q}^2
}\exp{\left[\frac{M_{\Omega^*_Q}^2-u_0\bar{u}_0m_\phi^2}{M^2}\right]}\left\{\frac{2u_0^2f_\phi
m_\phi
\left[\widetilde{\phi}_{\parallel}(\bar{u}_0)-\widetilde{g}_{\perp}^{(v)}(\bar{u}_0)\right]M^2E_0(x)}{\pi^2}\int_0^1 dt t\bar{t} e^{-\frac{\widetilde{m}_Q^2}{M^2}}\right.\nonumber\\
 &&- \frac{u_0^2f_\phi m_\phi m_Q^2
\left[\widetilde{\phi}_{\parallel}(\bar{u}_0)-\widetilde{g}_{\perp}^{(v)}(\bar{u}_0)\right]}{9M^4}
\langle\frac{\alpha_sGG}{\pi}\rangle\int_0^1
dt \frac{\bar{t}}{t^2} e^{-\frac{\widetilde{m}_Q^2}{M^2}}\nonumber\\
&&-\frac{u_0^2f_\phi m_\phi^3 \widetilde{A}(\bar{u}_0) }{2\pi^2}
\int_0^1 dt t e^{-\frac{\widetilde{m}_Q^2}{M^2}}+\frac{u_0^2f_\phi
m_\phi^3m_Q^2 \widetilde{A}(\bar{u}_0)
}{36M^6}\langle\frac{\alpha_sGG}{\pi}\rangle \int_0^1 dt
\frac{1}{t^2} e^{-\frac{\widetilde{m}_Q^2}{M^2}}\nonumber\\
&&+\frac{2u_0^3f_\phi m_\phi^3 \widetilde{\widetilde{C}}(\bar{u}_0)
}{\pi^2} \int_0^1 dt t\bar{t}
e^{-\frac{\widetilde{m}_Q^2}{M^2}}-\frac{u_0^3f_\phi m_\phi^3m_Q^2
\widetilde{\widetilde{C}}(\bar{u}_0)
}{9M^6}\langle\frac{\alpha_sGG}{\pi}\rangle \int_0^1 dt \frac{\bar{t}}{t^2} e^{-\frac{\widetilde{m}_Q^2}{M^2}}\nonumber\\
&&\left.  +\frac{u_0^2\widetilde{f}_\phi m_\phi
g_{\perp}^{a}(\bar{u}_0)}{72M^2}
\langle\frac{\alpha_sGG}{\pi}\rangle\int_0^1 dt
e^{-\frac{\widetilde{m}_Q^2}{M^2}}   \right\} \nonumber\\
&+&\frac{1}{\lambda_{\Omega^*_Q}^2}\exp{\left[\frac{M_{\Omega^*_Q}^2-m_Q^2-u_0\bar{u}_0m_\phi^2}{M^2}\right]}\left\{
\frac{4u_0^2\langle\bar{s}s\rangle f_\phi m_\phi m_s
\left[\widetilde{\phi}_{\parallel}(\bar{u}_0)-\widetilde{g}_{\perp}^{(v)}(\bar{u}_0)\right]}{3M^2}
 \right.\nonumber\\
  &&\left.+\frac{4u^3\langle\bar{s}s\rangle f_\phi m_\phi^3 m_s
\widetilde{\widetilde{C}}(\bar{u}_0)}{3M^4} \right\} \, .
 \end{eqnarray}

With the simple replacements,
 \begin{eqnarray}
\widetilde{ g}&\rightarrow& 2\widetilde{ g}, \, \,
\lambda_{\Omega^*_Q}^2\rightarrow\lambda_{\Omega^*_Q}
\lambda_{\Xi^*_Q}, \, \,M_{\Omega^*_Q}^2 \rightarrow
\frac{M^2_{\Omega^*_Q}+ M_{\Xi^*_Q}^2}{2}, \, \, f_\phi \rightarrow
f_{K^*}, \, \, f^\perp_\phi \rightarrow f^\perp_{K^*}, \, \, m_\phi
\rightarrow m_{K^*} ,\nonumber \\
\widetilde{ g}&\rightarrow& 2\widetilde{ g}, \, \,
\lambda_{\Omega^*_Q}^2\rightarrow\lambda_{\Sigma^*_Q}
\lambda_{\Xi^*_Q}, \, \,M_{\Omega^*_Q}^2 \rightarrow
\frac{M^2_{\Sigma^*_Q}+ M_{\Xi^*_Q}^2}{2}, \, \, f_\phi \rightarrow
f_{K^*}, \, \, f^\perp_\phi \rightarrow f^\perp_{K^*}, \, \, m_\phi
\rightarrow m_{K^*}, \nonumber \\
&&\langle\bar{s}s\rangle\rightarrow\langle\bar{q}q\rangle,
\,\,\langle\bar{s}g_s\sigma G
s\rangle\rightarrow\langle\bar{q}g_s\sigma G q\rangle, \,\, m_s
\rightarrow m_q,\nonumber \\
\widetilde{ g}&\rightarrow& 2\widetilde{ g}, \, \,
\lambda_{\Omega^*_Q}^2\rightarrow\lambda_{\Sigma^*_Q}^2 , \,
\,M_{\Omega^*_Q}^2 \rightarrow M^2_{\Sigma^*_Q}, \, \, f_\phi
\rightarrow f_{\rho}, \, \, f^\perp_\phi \rightarrow f^\perp_{\rho},
\, \, m_\phi
\rightarrow m_{\rho}, \, \,\langle\bar{s}s\rangle\rightarrow\langle\bar{q}q\rangle,\nonumber \\
&& \langle\bar{s}g_s\sigma G
s\rangle\rightarrow\langle\bar{q}g_s\sigma G q\rangle, \,\, m_s
\rightarrow m_q,
 \end{eqnarray}
we can obtain the corresponding strong coupling constants in the
vertexes $\Omega_Q^*\Xi_Q^*K^*$, $\Xi_Q^*\Sigma_Q^*K^*$,
$\Sigma_Q^*\Sigma_Q^*\rho$, respectively. Here $\bar{u}_0=1-u_0$,
$\widetilde{f}_\phi=f_\phi-f_\phi^\perp \frac{2m_s}{m_\phi}$,
$\widetilde{f}_\phi^\perp=f_\phi^\perp-f_\phi \frac{2m_s}{m_\phi}$,
$\widetilde{f}_{K^*}=f_{K^*}-f_{K^*}^\perp \frac{m_u+m_s}{m_{K^*}}$,
$\widetilde{f}_{K^*}^\perp=f_{K^*}^\perp-f_{K^*}
\frac{m_u+m_s}{m_{K^*}}$, $\widetilde{f}_\rho=f_\rho-f_\rho^\perp
\frac{m_u+m_d}{m_\rho}$,
$\widetilde{f}_\rho^\perp=f_\rho^\perp-f_\rho
\frac{m_u+m_d}{m_\rho}$,  $M_1^2=M_2^2=2M^2$ and
$u_0=\frac{M_1^2}{M_1^2+M_2^2}=\frac{1}{2}$ as
$\frac{M_{i}^2}{M_{i}^2+M_{j}}\approx \frac{1}{2}$,
 $\widetilde{m}_Q^2=\frac{m_Q^2}{t}$,
$E_n(x)=1-(1+x+\frac{x^2}{2!}+\cdots+\frac{x^n}{n!})e^{-x}$,
$x=\frac{s_0}{M^2}$;
$\widetilde{\widetilde{f}}(\bar{u}_0)=\int_0^{u_0}du\int_0^u dt
f(1-t)$, $\widetilde{f}(\bar{u}_0)=\int_0^{u_0}du f(1-u)$, the
$f(u)$ denote the light-cone distribution amplitudes, the lengthy
expressions of the light-cone distribution amplitudes
$\phi_{\parallel}(u)$, $\phi_{\perp}(u)$, $A(u)$, $A_\perp(u)$,
$g_\perp^{(v)}(u)$,
  $g_\perp^{(a)}(u)$, $h_{\parallel}^{(s)}(u)$,
  $h_{\parallel}^{(t)}(u)$, $h_3(u)$, $g_3(u)$, $C(u)
  $, $B_\perp(u)$, $C_\perp(u)$  can be found in
  Refs.\cite{VMLC2003,VMLC2007},


\begin{thebibliography}{99}
\bibitem{ReviewH1}  J. G. Koerner, D. Pirjol and M. Kraemer, Prog. Part. Nucl. Phys. {\bf 33} (1994) 787.

\bibitem{ReviewH2} F. Hussain, G. Thompson and J. G. Koerner, hep-ph/9311309.

\bibitem{H-baryon-1} E. Bagan, H. G. Dosch, P. Gosdzinsky, S. Narison and J. M. Richard, Z. Phys. {\bf C64} (1994) 57.

\bibitem{H-baryon-2} K. C. Bowler et al, Phys. Rev. {\bf D54} (1996)  3619.

\bibitem{H-baryon-3} E. Jenkins, Phys. Rev. {\bf D54} (1996) 4515.

\bibitem{H-baryon-4} S. S. Gershtein, V. V. Kiselev, A. K. Likhoded and A. I. Onishchenko, Mod. Phys. Lett. {\bf A14} (1999) 135.

\bibitem{H-baryon-5} D. Ebert, R. N. Faustov, V. O. Galkin, A. P. Martynenko, Phys. Rev. {\bf D66} (2002) 014008.

\bibitem{H-baryon-6} J. M. Flynn, F. Mescia and A. S. B. Tariq, JHEP {\bf 0307} (2003) 066.

\bibitem{H-baryon-7} J. Vijande, H. Garcilazo, A. Valcarce and F. Fernandez, Phys. Rev. {\bf D70} (2004) 054022.

\bibitem{H-baryon-8} A. Faessler et al, Phys. Rev. {\bf D73} (2006) 094013.

\bibitem{H-baryon-9} M. Karliner and H. J. Lipkin, Phys. Lett. {\bf B660} (2008) 539.

\bibitem{H-baryon-10} W. Roberts and M. Pervin, Int. J. Mod. Phys. {\bf A23} (2008) 2817.

\bibitem{H-baryon-11} T. M. Aliev, K. Azizi and A. Ozpineci, Phys. Rev. {\bf D79} (2009) 056005.

\bibitem{H-baryon-12} Z. G. Wang, Phys. Lett. {\bf B685} (2010) 59.

\bibitem{H-baryon-13} Z. G. Wang, Eur. Phys. J. {\bf C68} (2010) 479.

\bibitem{SVZ79} M. A. Shifman, A. I. Vainshtein and V. I. Zakharov, Nucl. Phys. {\bf B147} (1979) 385, 448.

\bibitem{PRT85} L. J. Reinders, H. Rubinstein and S. Yazaki, Phys. Rept. {\bf 127} (1985) 1.

\bibitem{LCSR89} I. I. Balitsky, V. M. Braun and A. V. Kolesnichenko, Nucl. Phys. {\bf B312} (1989) 509.

\bibitem{LCSR} V. L. Chernyak and A. R. Zhitnitsky, Phys. Rept. {\bf 112} (1984)173.

\bibitem{LCSRreview} P. Colangelo and A. Khodjamirian, hep-ph/0010175.

\bibitem{NarisonBook} S. Narison, Camb. Monogr. Part. Phys. Nucl. Phys. Cosmol. {\bf 17} (2002) 1.

\bibitem{Wang0909} Z. G. Wang, Phys. Rev. {\bf D81} (2010) 036002.

\bibitem{WangEPJA} Z. G. Wang, Eur. Phys. J. {\bf A44} (2010) 105.

\bibitem{Lee2010} S. H. Lee, A. Ozpineci and Y. Sarac, arXiv:1001.2905.

\bibitem{Azizi2009} K. Azizi, M. Bayar and A. Ozpineci, Phys. Rev. {\bf D79} (2009) 056002.

\bibitem{OsetPRL} V. K. Magas, E. Oset and  A. Ramos, Phys. Rev. Lett. {\bf 95} (2005) 052301.

\bibitem{Oset00} J. A. Oller, E. Oset and A. Ramos, Prog. Part. Nucl. Phys. {\bf 45} (2000) 157.

\bibitem{SU6} C. Garcia-Recio, J. Nieves and L. L. Salcedo, Phys. Rev. {\bf D74} (2006) 034025.

\bibitem{SU4-t-1} J. Hofmann and M. F. M. Lutz, Nucl. Phys. {\bf A763} (2005) 90.

\bibitem{SU4-t-2} J. Hofmann and M. F. M. Lutz, Nucl. Phys. {\bf A776} (2006) 17.

\bibitem{SU8} C. Garcia-Recio, V. K. Magas, T. Mizutani, J. Nieves, A. Ramos, L. L. Salcedo and L.
Tolos, Phys. Rev. {\bf D79} (2009) 054004.

\bibitem{SU4-t} C. E. Jimenez-Tejero, A. Ramos and I. Vidana, Phys. Rev. {\bf C80} (2009) 055206.

\bibitem{New-Baryon-1} J. L. Rosner, J. Phys. {\bf G34} (2007) S127.

\bibitem{New-Baryon-2} E. Klempt and J. M. Richard, Rev. Mod. Phys. {\bf 82} (2010) 1095.

\bibitem{Mizutani-2006} T. Mizutani and A. Ramos, Phys. Rev. {\bf C74} (2006) 065201.

\bibitem{Lutz-2006} M. F. M. Lutz and  C. L. Korpa, Phys. Lett. {\bf B633} (2006) 43.

\bibitem{V1} V. Pascalutsa, M. Vanderhaeghen and S. N. Yang, Phys. Rept. {\bf 437} (2007) 125.

\bibitem{V2} S. Nozawa and D. B. Leinweber,  Phys. Rev. {\bf D42} (1990) 3567.

\bibitem{Aliev0904}  T. M. Aliev, K. Azizi and M. Savci, Phys. Lett. {\bf B681} (2009) 240.

\bibitem{VMLC981} P. Ball,  V. M. Braun, Y. Koike and K. Tanaka,  Nucl. Phys. {\bf B529} (1998) 323.

\bibitem{VMLC982} P. Ball and V. M. Braun,  Nucl. Phys. {\bf B543} (1999) 201.

\bibitem{VMLC2003}  P. Ball and M. Boglione,  Phys. Rev. {\bf D68} (2003) 094006.

\bibitem{VMLC2007} P. Ball,  V. M. Braun and A. Lenz, JHEP {\bf 0708} (2007) 090.

\bibitem{Ioffe2005} B. L. Ioffe, Prog. Part. Nucl. Phys. {\bf 56} (2006) 232.

\bibitem{PDG} C. Amsler et al,  Phys. Lett. {\bf B667} (2008)  1.

\bibitem{Wang32} Z. G. Wang,  Eur. Phys. J. {\bf C68} (2010) 459.


\end{thebibliography}
\end{document}